\documentclass[11pt,twoside]{book}
\usepackage{konkolyproc2}
\usepackage{longtable}
\usepackage{amsmath,amssymb}
\usepackage{graphicx}
\usepackage{lscape}
\usepackage{index}
\usepackage{natbib}
\usepackage{bigdelim}
\usepackage{multirow}
\makeindex

\begin{document}

\pagestyle{myheadings}
\setcounter{equation}{0}\setcounter{figure}{0}\setcounter{footnote}{0}
\setcounter{section}{0}\setcounter{table}{0}\setcounter{page}{1}
\markboth{Jurkovic, Stojanovi\'c, \& Ninkovi\'c}{Galactic membership of BL Her 
type variable stars}
\title{Galactic membership of BL Her type variable stars}
\author{Monika I. Jurkovic, Milan Stojanovi\'c, \& Slobodan 
Ninkovi\'c}
\affil {Astronomical Observatory of Belgrade, Belgrade, Serbia}

\begin{abstract}
As the RR Lyrae stars evolve on the Hertzsprung-Russell 
diagram they are believed to become short period Type II 
Cepheids, known as BL Her type (with a pulsation period from 
$1$ to $3-8$ days). Assuming that their mass is around 
$0.5 - 0.6 {\rm M}_{\odot}$, and that they are low metallicity objects, 
they were thought to belong to the halo of the Milky Way. 
We investigated seven Galactic short period Type II Cepheids 
(BL Her, SW Tau, V553 Cen, DQ And, BD Cas, V383 Cyg, and KT Com) 
in order to establish their membership within the Galactic 
structure using the kinematic approach. $Gaia$ should provide 
us with more data needed to conduct the study of the whole sample.
\end{abstract}

\section{Introduction}

The evolution of low mass stars, such as RR Lyrae, can be followed up 
in the Type II Cepheids. \citet{Wallerstein2002} gives an overview of 
Type II Cepheids (T2C). The papers by \citet{Harris1984,Harris1985} 
discuss the classification of T2Cs according to 
their distances from the Galactic plane, while the papers by 
\citet{Diethelm1986,Diethelm1990} address the question of the relation of the 
metallicities and the position of these stars. By investigating the 
Galactic membership based on the kinematic approach we are able to 
reconstruct (within the limits of the model) the movement of an individual 
star in the Galaxy, which helps us to answer the question of the origin 
of metal-rich Type II Cepheids. The General Catalogue of Variable 
Stars\footnote{http://www.sai.msu.su/gcvs/gcvs/} (GCVS) in 2012 
contained 71 short period T2Cs, which were expanded in the time 
that has passed, but we stick to that sample, because they were 
relatively bright objects for which there was a chance to find all the 
data we needed (the distance or parallax, proper motion, and radial velocity).

\section{Method}

There are a few approaches for indicating the membership of stars of 
the Galactic components. 
Here, we shall use the kinematic approach. We start by converting the 
distance and position on the sky to the Galactocentric Cartesian 
system of Galactic coordinates $(X,Y,Z)$. For this we use the 
well-known formulas:

\begin{eqnarray}
X = D \cos b \cos l - R_{\odot}\nonumber\\
Y = D \cos b \sin l \\
Z = D \sin b \nonumber
\end{eqnarray}

In order to obtain the 3-D position vector and velocity vector of a 
star in space one needs the following data: two celestial coordinates, 
distance, two proper-motion components, and the radial velocity. 
Since these are pulsating variable stars they change their radial 
velocity due to pulsation too, so that one should be cautious when 
applying the radial velocity values.

In our original sample these necessary data were not available for 
all stars. For some stars we do not have the parallax or radial 
velocity, or both.  All the data necessary for calculating the velocity 
components are available for seven stars in total. All the input 
data should be transformed into the heliocentric Cartesian system; 
for this purpose we use the procedure described in 
\citet{JohnsonSoberblom1987}. Then we correct velocity values for 
the solar motion. The obtained velocity components $U, V, W$ are 
with respect to the local standard of rest (LSR).

The magnitude of the LSR velocity $v$, 

\begin{equation}
v=\sqrt{U_{\rm LSR}^2 + V_{\rm LSR}^2 + W_{\rm LSR}^2}
\end{equation}

\noindent is indicative of the star membership, to the thin disc, 
thick disc, or halo. If for a star the magnitude of the LSR velocity 
is very high (say, exceeds 250~km\,s$^{-1}$), then the probability 
that this star belongs to the thin or thick disc is very low. 
If it exceeds, say 100~km\,s$^{-1}$, then only the probability of 
belonging to the thin disc is very low.

\section{Results}

The stars which had all the required data are listed in 
Table~\ref{table1:stars}.

\begin{table}[ht]
\caption{Stars examined for their Galactic membership (in the order 
of increasing period).}
\label{table1:stars}
\smallskip
\begin{center}
\scriptsize
\begin{tabular}{lccr@{\hskip 2mm}rcc}
\tableline
\noalign {\smallskip} 
Name & RA (J2000) & Dec (J2000) & Proper & motion\ \  & Radial velocity & Parallax \\
& [h : m : s] & [$^\circ$ : $^\prime$ : $^{\prime\prime}$] & [mas/yr] & [mas/yr] & [km/s] & [mas]\\
\noalign{\smallskip}
\tableline
\noalign{\smallskip}
BL Her   & 18:01:09.22 & +19:14:56.68 & $-$2.94 & $-$12.94 & 18.0 & 1.27\\
SW Tau   & 04:24:32.97 & +04:07:24.05 &    4.05 & $-$11.17 & 10.9 & 2.8\\
V553 Cen & 14:46:33.63 & $-$32:10:15.25 &  5.01 & $-$0.71 & $-$6.00 & 1.84\\
DQ And   & 00:59:34.47 & +45:24:24.22 &    5.16 &    1.92 & $-$230.91 & 0.67\\
BD Cas   & 00:09:51.39 & +61:30:50.54 & $-$1.1 &  $-$0.9 & $-$49.30 & 2.13\\
V383 Cyg & 20:28:58.15 & +34:08:06.36 & $-$1.99 & $-$2.64 & $-$24.4 & 4.44\\
KT Com   & 13:33:50.22 & +17:25:30.37 & $-$15.93 & $-$24.76 & $-$13.0 & 5.50\\
\noalign{\smallskip}
\tableline
\noalign{\smallskip}
\end{tabular}  \end{center}
\scriptsize
 Note: All the data were collected from: http://simbad.u-strasbg.fr/simbad/, 
 and from the references within \citet{vanLeeuwen2002} for the Hipparcos data. 
 \end{table}
 
\normalsize

\begin{figure}[!ht]
\includegraphics[width=1.0\textwidth]{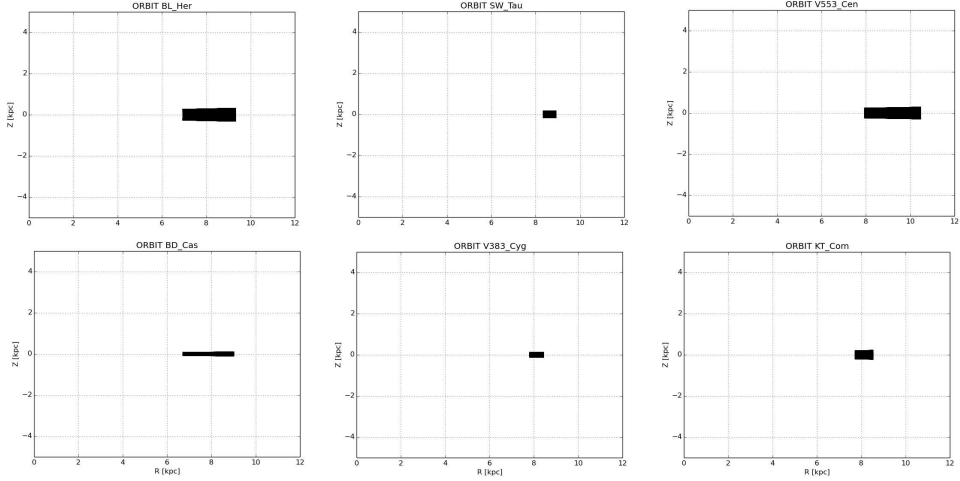}
\caption{The meridional plots of star orbits for: BL Her, SW Tau, 
V553~Cen, BD Cas, V383 Cyg and KT Com in the order of their increasing 
periods. $R$ is the distance from the Galactic rotation axis, $z$ is 
the distance from the Galactic plane.} 
\label{jurkovic-fig1} 
\end{figure}

\begin{figure}[!tbp]
  \centering
  \begin{minipage}[b]{0.45\textwidth}
    \includegraphics[width=\textwidth]{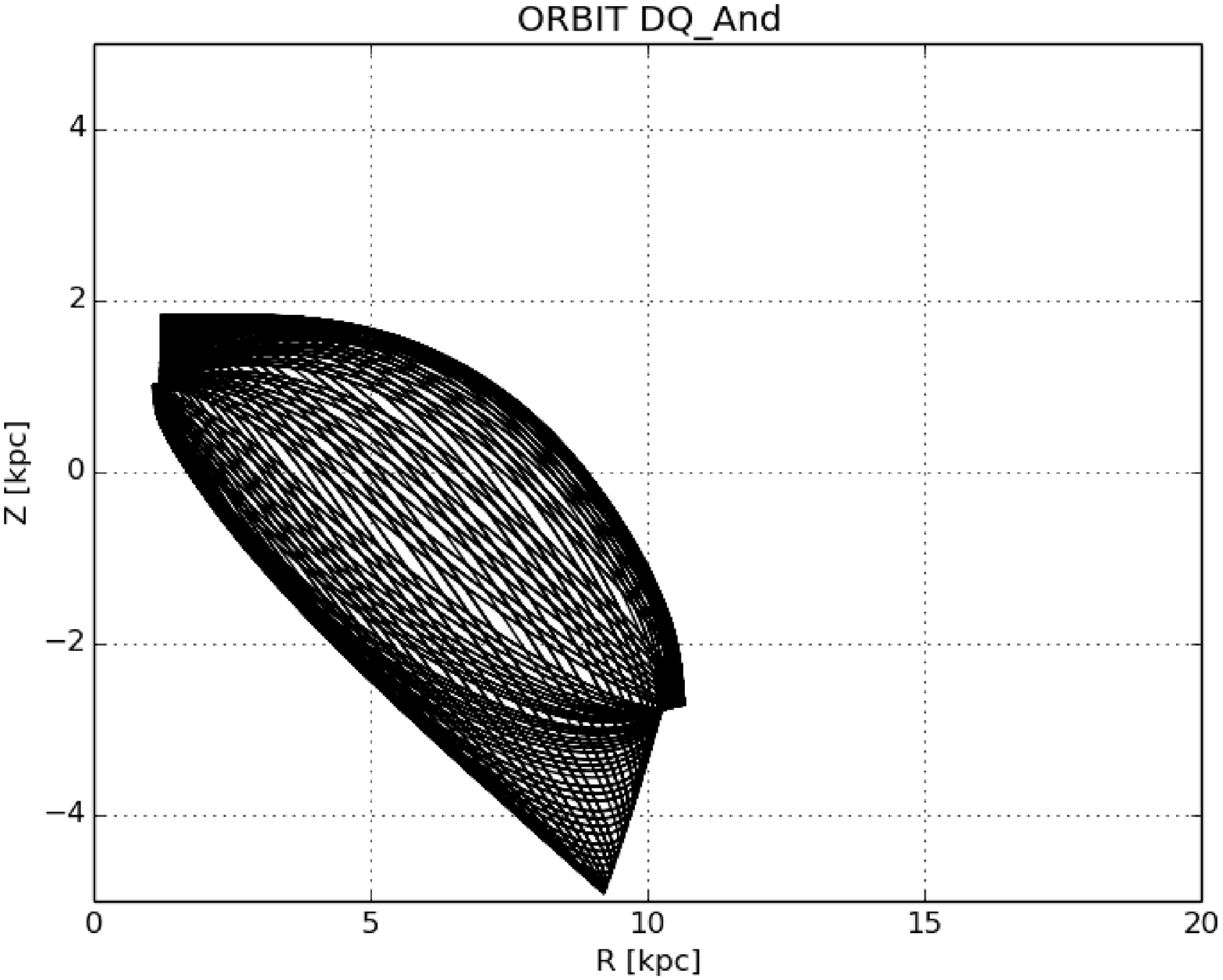}
    \caption{The meridional plot of DQ And. $R$ and $z$ are the same 
    as in Figure~\ref{jurkovic-fig1}.}
    \label{jurkovic-fig2}
  \end{minipage}
  \hfill
  \begin{minipage}[b]{0.5\textwidth}
    \includegraphics[width=\textwidth]{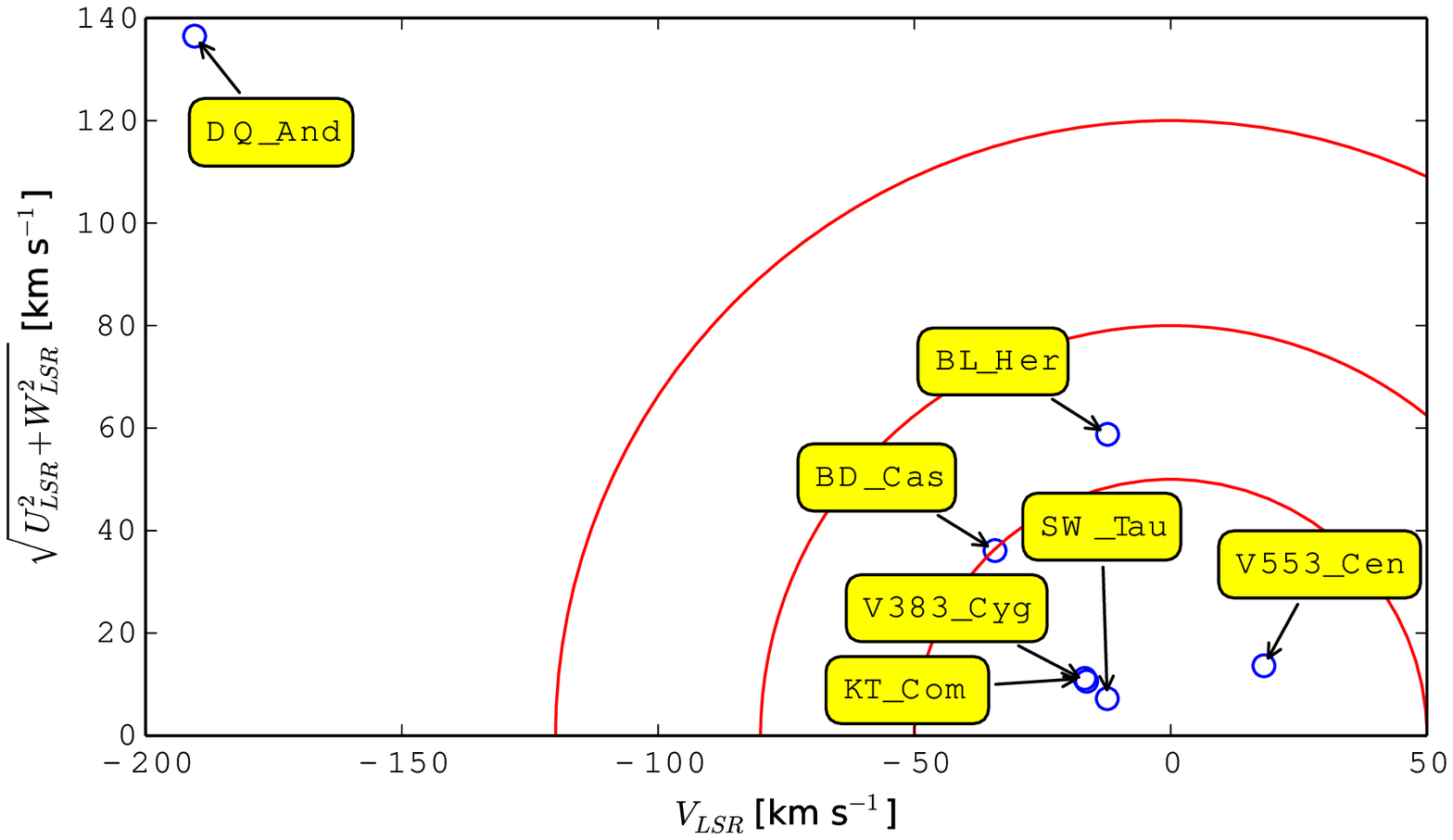}
    \caption{The Toomre diagram of the examined stars.}
    \label{jurkovic-fig3}
  \end{minipage}
\end{figure}

Figures~\ref{jurkovic-fig1} and \ref{jurkovic-fig2} show the 
cross-section of the path of each stars in our sample in the past 12 Gyrs from the 
model calculations. All the stars in Figure~\ref{jurkovic-fig1} are 
members of the thin disc. DQ Andromedae (see Figure~\ref{jurkovic-fig2}) 
is a halo star, but with a mean metallicity of [Fe/H] = $-0.17$ 
\citep{Schmidt2011}. Since the distances to almost all the stars 
are probably not precise enough, the results could change, but not too much.

The Toomre diagram in Figure~\ref{jurkovic-fig3} shows the 
distribution of the examined stars in the calculated velocity 
planes. The lines show the approximate limits between the subsystems 
in the Galaxy: the innermost part being the thin disc, then the 
thick disc, and finally the halo.

\section{Conclusion}

Out of the 71 stars we have studied, only 7 had enough data 
available to be examined in our model. Even though we are aware 
of a substantial error influence in the input data, we are still 
comfortable with stating that the model results do give us the 
Galactic membership of the stars.

Kinematically BL Her, SW Tau, V553 Cen, BD Cas, V383 Cyg, and KT~Com 
could be thin disk stars, but by examining their light curve shapes it 
might happen that they turn out to be some other type of variable 
stars, not Type II Cepheids. 

DQ And is the only star which shows evidence of being a member of 
the halo of the Milky Way, but the asymmetry of its calculated 
orbit is peculiar. If we consider that the Milky Way has experienced 
collisions with neighbouring dwarf galaxies, this asymmetry could 
be due to the capture of this star by our Galaxy or its motion 
around the centre of the Milky Way might have been perturbed. 

In the following years the astrometric measurements from the $Gaia$ 
satellite should give us much more insight into the understanding 
of the Galactic membership of Type II Cepheids.

\section*{Acknowledgements}

The authors thank for the financial support from the Ministry 
of Education, Science and Technological Development of the 
Republic of Serbia through the projects 176004 and 176011 and 
the Hungarian National Research, Development and Innovation Office through 
NKFIH K-115709. 
This research has made use of NASA's Astrophysics Data System. 
This research has made use of the VizieR catalogue access tool, 
CDS, Strasbourg, France. The original description of the VizieR 
service was published in \citet{VizieR}.

\end{document}